# Keep on growing or change direction? Dependence of clonal plant survival on the probability of side-branching

Róbert Juhász[1*] and Beáta Oborny[2,3,4,*]

1. HUN-REN Wigner Research Centre for Physics, Institute for Solid State Physics and Optics,
H-1525 Budapest, P.O.Box 49, Hungary.
Email: juhasz.robert@wigner.hun-ren.hu

2. Institute of Botany, Charles University,
Benátská 2, Prague 2, 12800, Czech Republic

3. Institute of Evolution, HUN-REN Centre for Ecological Research,
H-1121 Budapest, Konkoly-Thege M. Road 29-33, Hungary

4. Institute of Biology, Eötvös Loránd University,
H-1117 Budapest, Pázmány P. stny. 1C, Hungary.
Email: beata.oborny@ttk.elte.hu

* The authors contributed equally to this work.

**Corresponding author:**
Beáta Oborny, email: beata.oborny@ttk.elte.hu

## Abstract

In clonal organisms, the genetic individual (genet) may consist of several physiological individuals (ramets). Clonal plants typically grow horizontally by means of stolons or other clonal growth organs. Some species grow linearly for several ramet generations, while others produce side branches frequently. We hypothesize that the probability of side-branching (*b*) affects genet survival (*G*) depending on the ramets' ratio of birth to death rates (λ). We used a spatially explicit model of ramet population dynamics to compare three growth strategies. In the neutral case, growing forward and side-branching were equally probable, whereas forward-growing and side-branching plants had lower and higher *b*, respectively. We found that the genet's survival threshold (the critical $\lambda_c$) strongly depended on the branching probability. The lowest threshold, and hence the broadest range of survival, occurred in moderately forward-growing plants, in which the average length of linear (non-branching) intervals was 1.89 units. Above the threshold, the *G(b)* relationship strongly depended on λ. Genet survival was promoted by forward growth in sparse ramet populations (low λ). However, this advantage gradually disappeared as λ increased, and branching probability became effectively neutral across a broad parameter range encompassing both forward-growing and side-branching plants. This suggests strong selection for low branching intensity in unfavorable habitats, but weak or negligible selection in favorable habitats.

# 1. Introduction

The study of spatial population dynamics and pattern formation has long been an important area of research in mathematical ecology. Modelling clonal plants is particularly interesting because a genetic individual (genet) can consist of multiple physiological individuals (ramets), each potentially possessing all the organs required for independent life. Before becoming independent, an offspring ramet can be affected by its parent in multiple ways. The parent can influence the offspring's position and affect its development through the transfer of resources and hormones (Oborny et al., 2012).

The genet typically grows horizontally by means of stolons or other clonal growth organs (Klimešová et al., 2017). New stolons are typically added to the genet at discrete points, where apical or lateral meristems are available (**Figure 1**). Biomass allocation to apical versus lateral growth is an important developmental decision (Huber et al., 1999), as it affects the encounter probabilities of branches (Bell and Tomlinson, 1980), and, consequently, self-inhibition within the clone. The encounter probabilities are also affected by the birth and death rates of ramets (Oborny and Marcsó, 2023). Altogether, a plant clone can be viewed as a complex network, in which the nodes (ramets) can affect one another through their links (stolons or other clonal growth organs; Oborny 2019).

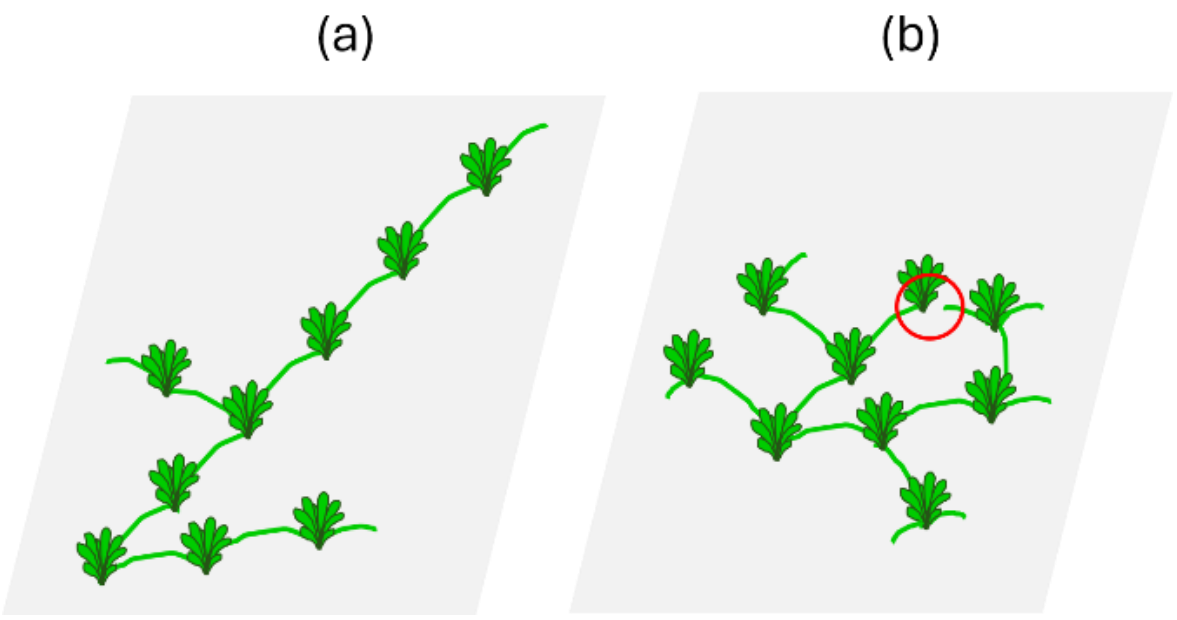

**Figure 1.** Schematic examples of clonal plants growing on continuous planes (grey). Each genet consists of nine ramets connected by stolons. The branching probability is lower in **(a)** than in **(b)**. Open-ended lines represent growing stolon tips. The red circle highlights a case in which a stolon is attempting to colonize a site already occupied by a ramet.

The present study focuses on a simple parental influence: control of the offspring's direction relative to the parent. This control occurs in almost all clonally growing species[1], but varies in strength and preferred direction. There are species in which forward growth is strongly preferred, with growth continuing growth along the grandparent-parent principal axis (**Figure 1.a**). For example, *Carex arenaria* produces long, linear rhizomes with infrequent branching (Bell and Tomlinson, 1980). In this species, linear sequences comprising up to 87 ramets have been reported, while other extended up to 12 meters (D'Hertefeldt and Jónsdóttir, 1999). Other examples of such forward-growing species include *Psammochloa villosa* (Yu et al., 2004), and *Ipomoea pes-caprae* (Castellani and Santos, 2006). Conversely, there are species that prefer sideways growth, allocating biomass primarily to new lateral branches (**Figure 1.b**). For example, *Solidago canadensis* was observed to place only 18% of its new ramets in the principal direction, with most ramets developing at approximately ±60° to either side of it (Smith and Palmer, 1976). In more extreme cases, the plant can completely abandon growth in the principal direction and grow exclusively sideways. A well-known example is *Alpinia speciosa*, which branches at approximately ±60° to the left and right, without any forward growth, producing a hexagonal branching pattern (Bell and Tomlinson, 1980). The rules of branching have been described for many species within the framework of plant architecture[2] (Barthélémy and Caraglio, 2007; Bell, 1984). The above-mentioned rules may be rigid or phenotypically plastic (Alpert and Simms, 2002; Hutchings and de Kroon, 1994; Oborny et al., 2012). For example, auxin production by the apex can strongly suppress lateral branching, thereby promoting forward growth (Cline, 1997).

[1] We restrict our study to species in which ramets are anchored in space (e.g., by roots). Freely dispersing types of clonal offspring (e.g., bulbils) are not included. Concerning dispersal, they are more similar to seeds, and thus fit into the scope of the standard CP model (see below).

Within this diversity of clonal plant architectures, we focus on biomass allocation to forward growth versus side-branching. We investigate the effect of the probability of side-branching ($b$) on genet survival ($G$). We do not investigate the branching angle, because previous studies have shown that it has little effect on genet performance (Benedek et al., 2017; Sutherland and Stillman, 1988). We use a simple lattice model of clonal growth to demonstrate some basic phenomena. We show that genet survival is not a strictly monotonic function of branching probability and that the shape of this $G(b)$ function depends on the ratio of ramet birth to death rates ($\lambda$).

[2] We emphasize side-branching throughout the text, because forward growth is achieved by new branches in the case of sympodial clonal growth organs.

## 2. Methods

### *2.1. A new model type: the directional contact process (DCP)*

Our model is based on the contact process (CP, Harris, 1974), a paradigmatic model of spatial population dynamics. This is a spatially explicit version of the well-known Levins model of space occupation (Levins, 1969). The study of clonal plants necessitated an extension of the standard contact process. In the standard CP, the neighboring cells are attenpted to be colonized with equal probabilites (**Figure 2.a**). This can adequately represent the spread of non-clonal plants through seed dispersal, whereby the direction of colonization is not actively determined by the parent. In contrast, stolon growth in clonal plants is an active process governed by architectural rules. The direction of colonization can depend on the direction of the preceding colonization step. We modified the standard CP accordingly, introducing the 'directional contact process' (DCP), in which the colonization probabilities of the neighboring sites are not equal but depend on the direction of the preceding colonization step (**Figure 2.b–d**).

Although the contact process has been thoroughly studied with many kinds of modifications (primarily in statistical physics; Hinrichsen, 2000; Marro and Dickman, 1999; Ódor, 2004), we are not aware of any previous model with the modification described above. Note that directionality is local in the present case, taking into account only the preceding colonization of the parent site. This contrasts with a globally directional model, in which colonization has a uniform preferred direction throughout the system (as, for example, in a one-dimensional model by Tretyakov et al., 1997).

First, we examined whether the population dynamics of non-clonal individuals in the standard CP differs from the population dynamics of clonal ramets in the DCP. Secondly, we compared three clonal growth strategies within the DCP: the neutral, forward-growing, and side-branching strategies, which differed in their values of $b$ (**Figure 2.b–d**). We examined the broadest possible range of $b$ at selected values of the demographic parameter $\lambda$.

### *2.2. Model details*

The habitat was represented by a two-dimensional square lattice with a four-cell neighborhood (**Figure 2**). Each lattice cell could host a single ramet, and the lattice constant corresponded to the dispersal (parent-to-offspring) distance.

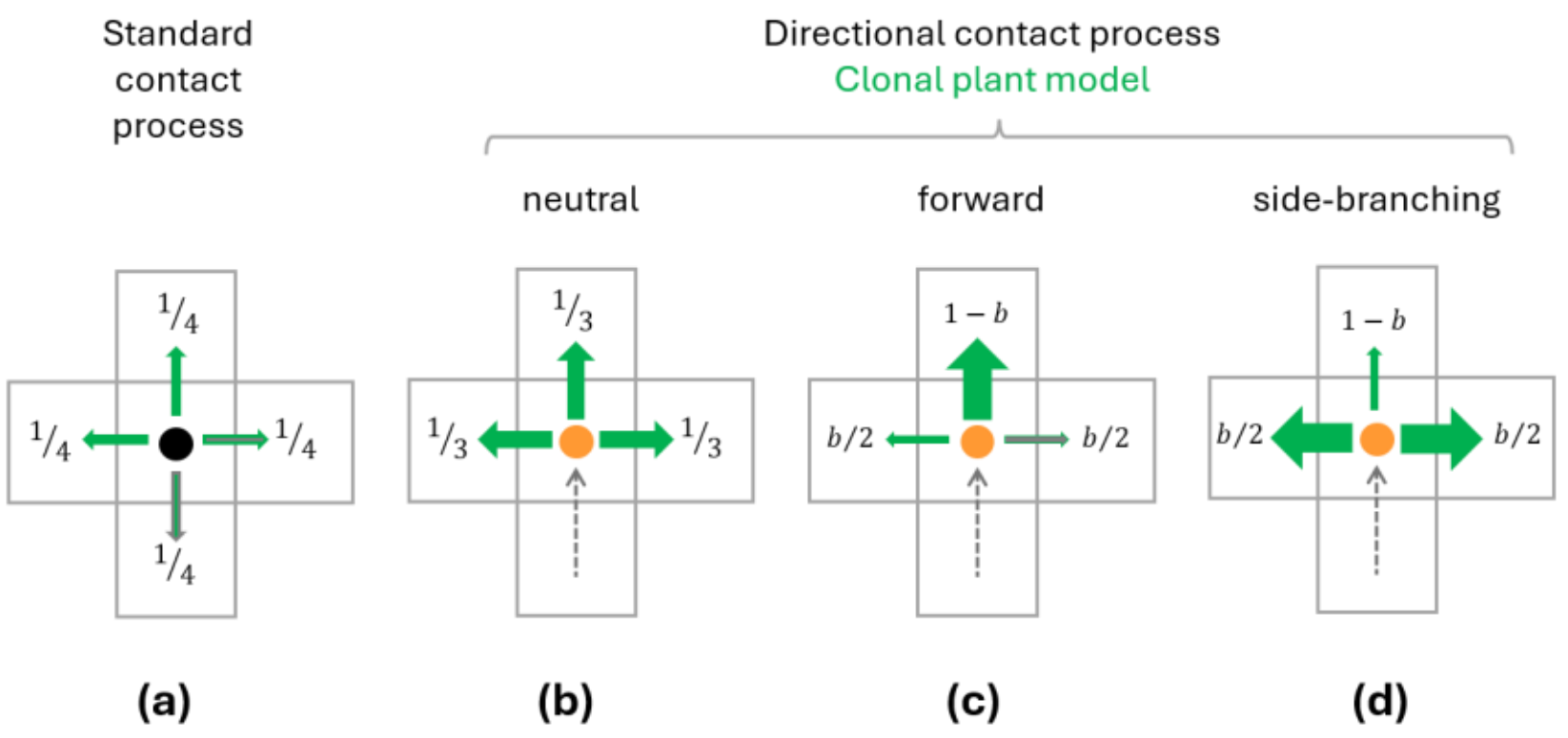

**Figure 2.** The standard contact process and its extension, developed to represent clonal plant growth. Each panel depicts a selected (parental) cell on a square lattice, from which colonization can occur into neighboring cells. A black dot indicates an individual of a non-clonal plant, while orange dots represent ramets of a clonal plant. Grey dashed arrows indicate the preceding colonization event, from the grandparent to the parent. Solid arrows show the potential directions of colonization. The probabilities of attempting colonization are shown in the cells and indicated by the widths of the solid green arrows. **(a) Standard CP:** all directions are equally probable. **(b) Neutral growth:** turning back to the grandparent's site is impossible, but growth in the remaining three directions are equally probable ($b = 1/3 + 1/3 = 2/3$). **(c)** Forward growth: continuation in the grandparent-parent direction is preferred ($b < 2/3$). **(d) Side-branching growth:** the other two directions are preferred ($b > 2/3$).

In the standard CP, the state of each lattice site is customarily characterized by a binary variable: empty vs. occupied (in other terms, inactive vs. active). In our DCP, occupied sites could be in one of four states ($s_i$=1, 2, 3, or 4) encoding the direction (North, East, South, or West) from which the parent colonized the site. Thus, together with the empty state ($s_i$=0), each site can be in five different states.

When a ramet died, the occupied cell became empty (local extinction), whereas in the event of ramet birth, an empty cell became occupied (colonization). The extinction probability was independent of the neighborhood and was uniform throughout the lattice. This feature is shared by the standard CP and the DCP. The rules of colonization, however, were different. In the standard CP, all cells in the parent's neighborhood have an equal probability (1/4) of being selected for colonization (**Figure 2.a**), whereas in the DCP, we took the grandparent-parent direction into account. As a general rule, it was impossible to turn back into the grandparent's site. Therefore, colonization of the grandparent's site could not occur immediately, but only through multiple growth steps. The types of DCPs differed in the probabilities of colonization attempts in the remaining three cells (**Figures 2.b–d**). In general, the probability of selecting the principal (forward) direction was *1-b*, whereas the total probability of selecting either side-branching direction was *b*. The left or right direction was selected at random, with equal probabilities (1/2). The value of *b* ranged between $0 \leq b \leq 1$. In the extreme case *b*=0, no side-branching occurred, thus the growing clone was one-dimensional. At the other extreme, *b*=1, the principal direction was always abandoned and growth continued only in the side-branches. **Figure 1** shows two intermediate cases with low versus high values of *b*.

### *2.3. Numerical simulations*

Each simulation represented the growth of a single genet. The dynamics of the genet's space occupation was implemented as follows. Within each elementary step, an occupied site was selected from the list of all occupied sites randomly, with equal probabilities. Then one of the following two actions was carried out. 1) With probability *e*, the selected site became empty (state 0), i.e., ramet death occurred. 2) With the complementary probability *c=1-e*, a single reproduction event was attempted. The direction of the attempt was selected according to the probabilities shown in **Figure 2**. If the selected target site was empty, then it became occupied, that is, a ramet birth occurred; otherwise, nothing happened. In the standard CP, the occupied target site's state was coded as 1, whereas in the DCP, we recorded the direction of colonization (North, East, South, or West, as indicated by the dashed arrow in **Figure 2**). After the elementary step, time *t* was incremented by $\Delta t = 1/N\ (t)$, where *N(t)* was the number of ramets, i.e., the length of the list of occupied sites, at time *t*. Thus each occupied site was updated once on average per unit time. Using this updating rule, the dynamics described above can be regarded as an embedded Markov chain of a continuous-time Markov process with transition rates *e* and *c=1-e*. The difference from the continuous-time process is that the time increment is not stochastic but is fixed at its mean value.

First, we studied the dependence of genet survival on branching probability *G(b)*, at various values of the demographic parameter $\lambda = c/e$. We used a single-seed initial condition, in which a central cell of the lattice was set to one of the four occupied states (the choice was irrelevant), while all other cells were empty. At $\lambda > 2$, the lattice size was 1,000 x 1,000 cells and the simulation time was $t \leq 2^{10}$; whereas at $\lambda \leq 2$, the lattice size was 10,000 x 10,000 cells and the simulation time $t \leq 2^{16}$. During these simulation times, the genet did not reach the boundaries of the lattice, thus finite-size effects were excluded. The durations of the simulations were were sufficient for $G(b, \lambda)$ to reach a steady value. The number of independent repetitions was $10^5$ at each parameter combination. Secondly, we focused on estimating the survival threshold, i.e., the lowest $\lambda$ value above which $G(b, \lambda) > 0$. In these simulations, the lattice size was 10,000 x 10,000 lattice cells, the duration was $t \leq 2^{16}$, and the number of independent repetitions was $10^5$ at each parameter combination. In the next section, we explain in detail how the survival threshold was estimated.

*2.4. Characterizing extinction as a critical transition*

Earlier investigations of the standard CP have shown that extinction of the entire population can be considered as a critical transition (Hinrichsen, 2000). We briefly review those features of this transition that are relevant for our present research. Importantly, the steady-state ramet population density $\hat{n}$ depends only on the ratio $\lambda = c/e$. Therefore, $\lambda$ is a suitable control parameter of the transition. We studied the genet's survival probability $G(b, \lambda)$ across the entire range $0 \leq b \leq 1$, in fine resolution (0.05), at six selected values of $\lambda$.

At sufficiently high values of $\lambda$, the steady-state ramet population density is non-zero, $\hat{n} > 0$. However, as $\lambda$ decreases, the system reaches a threshold at and below which $\hat{n}$ becomes zero. The same holds for the survival probability *G*. This transition from the living to the extinct state is a critical transition (a second-order phase transition) that obeys characteristic scaling laws near the threshold (Hinrichsen, 2000). The critical threshold has been estimated $\lambda_c \cong 1.6448(1)$ in the standard CP in the limit of infinite system size (Marro and Dickman, 1999; see Oborny et al., 2007 for more details about the standard CP from an ecological point of view). The number in brackets denotes the error of the estimate. We estimated $\lambda_c$ in the three versions of the DCP, providing a novel result.

The estimation of $\lambda_c$ is not a trivial task, because the system is known to show a 'critical slowing down' in the vicinity of the threshold. That is, the time required to reach $\hat{n}$ goes to infinity as $\lambda \rightarrow \lambda_c$ from above in an infinite system. Consequently, the measurement of $G(\lambda)$ would require very long simulation times. We applied another method of estimating $\lambda_c$ that is standard in statistical physics and utilizes one of the scaling laws that is valid for large times,

$$G(t) \sim t^{-\delta}, \qquad \text{Eq. 1}$$

where *t* is time, and $\delta$ is a critical exponent. This relationship applies strictly at the critical threshold, as *G(t)* tends exponentially to its limiting value above and below the threshold (Marro and Dickman, 1999). The critical threshold was detected by plotting *G(t)* on double-logarithmic scale. In this case, $G(t)t^{-\delta}$ is a straight line with a slope $-\delta$, whereas it is convex above the threshold and concave below it. Using this property of *G(t)*, we determined the critical control parameter by successive approximation up to four digits for a series of *b*.

It is important to note that the standard CP is a representative of an entire universality class of critical phenomena (the DP universality class; Ódor, 2004). Thus, Eq. 1 is applicable not only to a single model but to an entire class of models, which includes the spatial Levins model and spatial logistic model of population dynamics (Oborny et al., 2007). According to the DP conjecture, we also expect the DCP model to fall into this universality class, which is confirmed by our numerical results. For all two-dimensional models within this universality class, not only are the scaling relations identical in form (Eq. 1), but the value of the scaling exponent, $\delta = 0.4505(10)$, is also the same (Voigt, 1997). Thus our qualitative results do not depend on the geometry of the lattice, or on the (finite) neighborhood size. Even continuous space could be assumed. $\delta$ depends only on the dimensionality of the system (Hinrichsen, 2000).

## 3. Results

### *3.1. The standard vs. directional contact process*

In the standard CP, the four neighboring cells were selected for colonization with equal probability (1/4; **Figure 2.a**). The survival threshold in the limit of infinite system size had already been published, $\lambda_c \cong 1.6448(1)$ (Marro and Dickman, 1999). The purple curve in **Figure 3** shows the survival probability, *G(λ)*, over a broad range of $\lambda$. The curve demonstrates the abrupt decline of survival probability in the vicinity of the survival threshold (c.f. Marro and Dickman, 1999).

We compared the directional CP (DCP) with this standard model. The green curve in **Figure 3** shows the neutral DCP, in which backtracking to the grandparent's site was not possible and the forward, left, and right directions were equally probable (**Figure 2.b**). The survival threshold proved to be lower than in the standard CP, and the survival probability *G(λ)* was generally higher. Therefore, prohibiting backtracking to the grandparent's site facilitated survival. This is understandable, as the grandparent's site may still be occupied by the grandparent when the offspring arrives. However, the light blue curve in **Figure 3** demonstrates that the probability of side-branching (*b*) should also be taken into account. In that example, with *b*=0.1, survival was lower in the DCP than in the standard CP. The next section analyses $\lambda_c(b)$ in more detail.

**Figure 3.** Dependence of genet survival probability (*G*) on the demographic rate of ramets (λ). The curves show the standard CP, the neutral DCP (*b*=2/3), and a non-neutral DCP representing a strongly forward-growing plant (*b*=0.1).

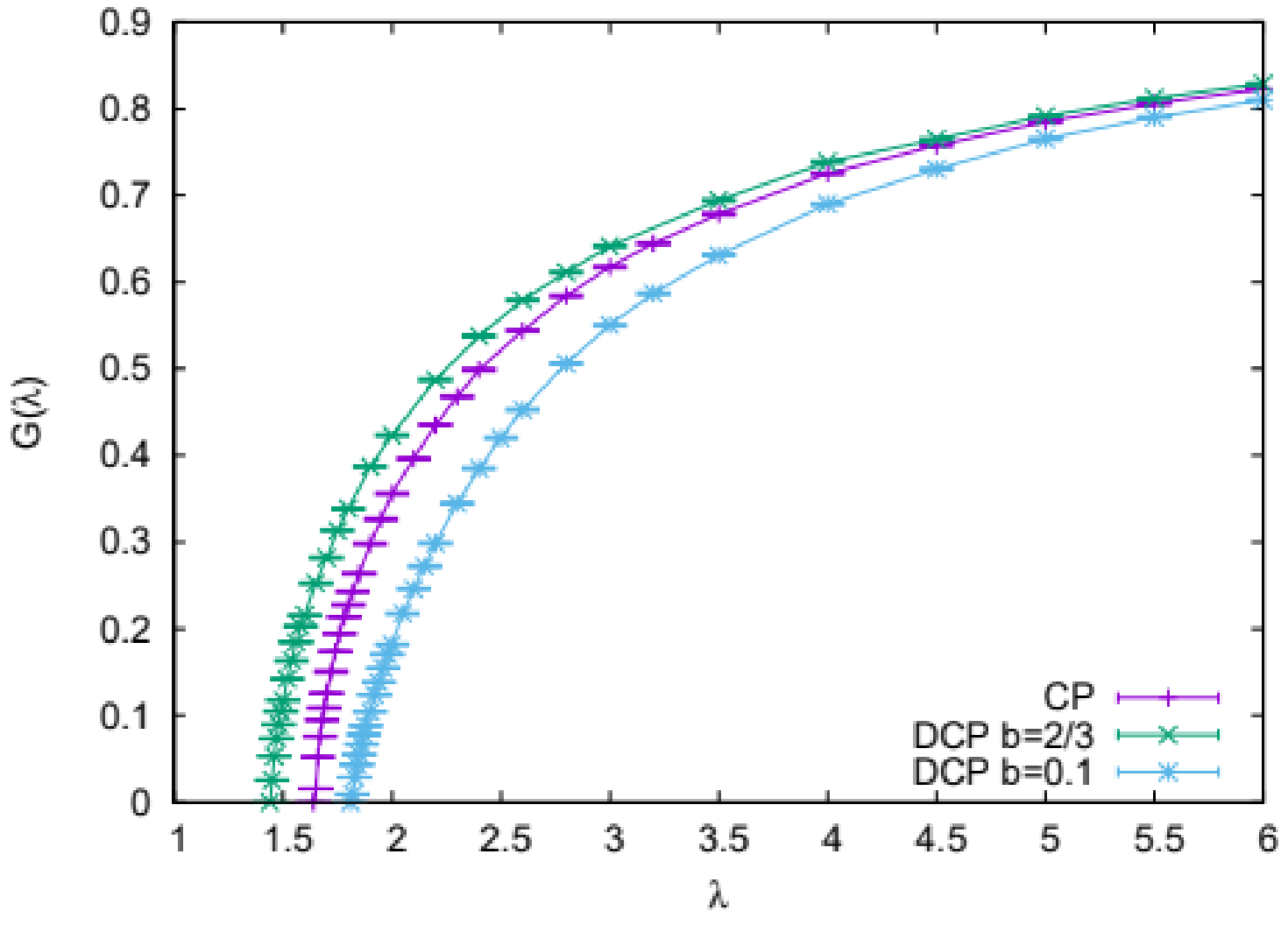

3.2. *The effect of side-branching on the survival threshold*

**Figure 4** presents the effect of *b* across its full range, [0,1]. The difference between the red dot (at *b*=2/3) and the dotted line shows, as in **Figure 3**, that preventing backtracking is sufficient to significantly improve genet survival. However, survival can be further improved by a moderate tendency for forward growth. The minimum of the $\lambda_c(b)$ curve is at approximately *b*=0.53(1), when the average length of linear (i.e., non-branching) segments is 1/*b* = 1.89(4) lattice units. The number in brackets denotes the error of the estimate. In comparison, this length is 1.5 lattice units in the case of neutral growth. These values depend on the details of the model (lattice geometry and neighbourhood size), but generally indicate the advantage of slight forward growth. The range of *b* in which this advantage over the neutral case was observed was quite broad, $0.4 \leq b < 2/3$.

Any stronger tendency for forward growth proved to be disadvantageous, compared to neutral growth. Furthermore, the disadvantage was evident even relative to the standard CP for $b < 0.2$. At the limit, *b*=0.2, the linear segments were quite long, having an average of 1/*b*=5 lattice units. Finally, at *b*=0, the plant did not branch at all, and its survival threshold became equal to that of the one-dimensional, one-sided contact process with $\lambda_c(b=0)$= 3.306(2) (Tretyakov et al., 1997). We exclude this special one-dimensional case from the further analyses.

**Figure 4.** The effect of the side-branching probability (*b*) on the survival threshold ($\lambda_c$). The red circle highlights the neutral case (*b*=2/3). Blue circles to the left of this point (*b*<2/3) show forward-growing plants, while those on the right (*b*>2/3) show side-branching plants. For reference, the dotted line shows the survival threshold of the standard CP ($\lambda_c \cong 1.6448$).

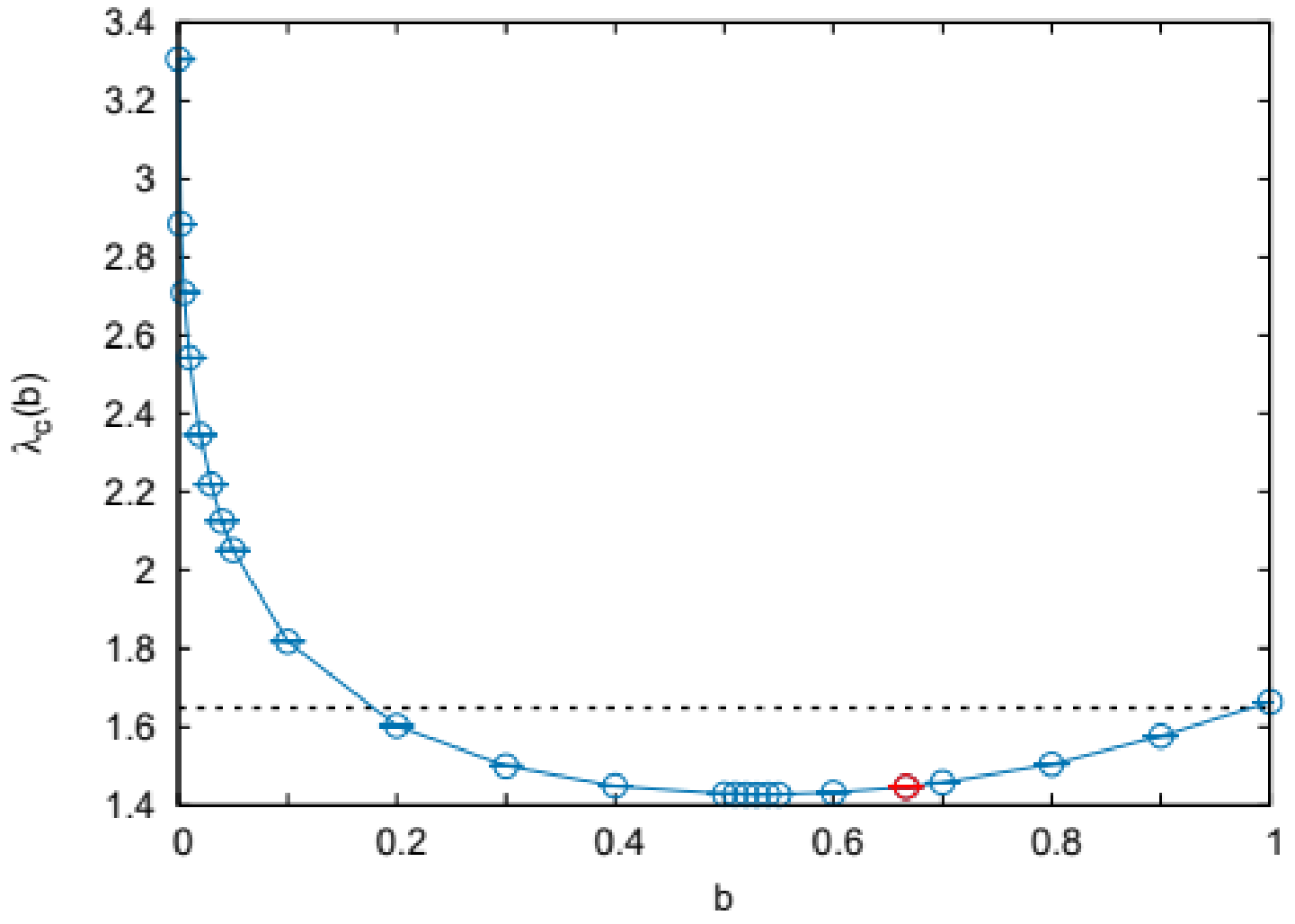

Intensive side-branching (*b*>2/3) was less favourable than neutral growth, but only the complete lack of forward growth (*b*=1) caused the survival threshold approach that of the standard CP, slightly exceeding it. This demonstrates that the lack of backtracking is sufficient to improve survival, even when combined with weak forward growth.

3.3. *The effect of side-branching on the survival probabilities above the threshold*

In the previous section, we studied the behavior of the system at the survival threshold, where $G \to 0$. Now we present its behavior above the threshold, over a broad range of $\lambda$.

The lowest, purple curve in **Figure 5** shows *G(b)* when the system is below the extinction threshold of the standard CP ($\lambda < \lambda_c \cong 1.6448$). As we can see in **Figure 4**, the plant was still viable at this $\lambda$=1.5, but not across the entire range of *b*. Only intermediate branching probabilities allowed genet survival. A moderate tendency for forward growth was more advantageous than neutral growth (that is, the curve is above the red dot) over the range $0.5 \leq b < 2/3$. Increasing the value of $\lambda$, the probability of survival increased monotonically. The range of *b* permitting survival expanded, and the curve flattened. This suggests that branching probability was gradually becoming a neutral trait.

**Figure 5.** The effect of the side-branching probability (*b*) on the probability of genet survival (*G*). Red dots mark the neutral case (*b*=2/3).

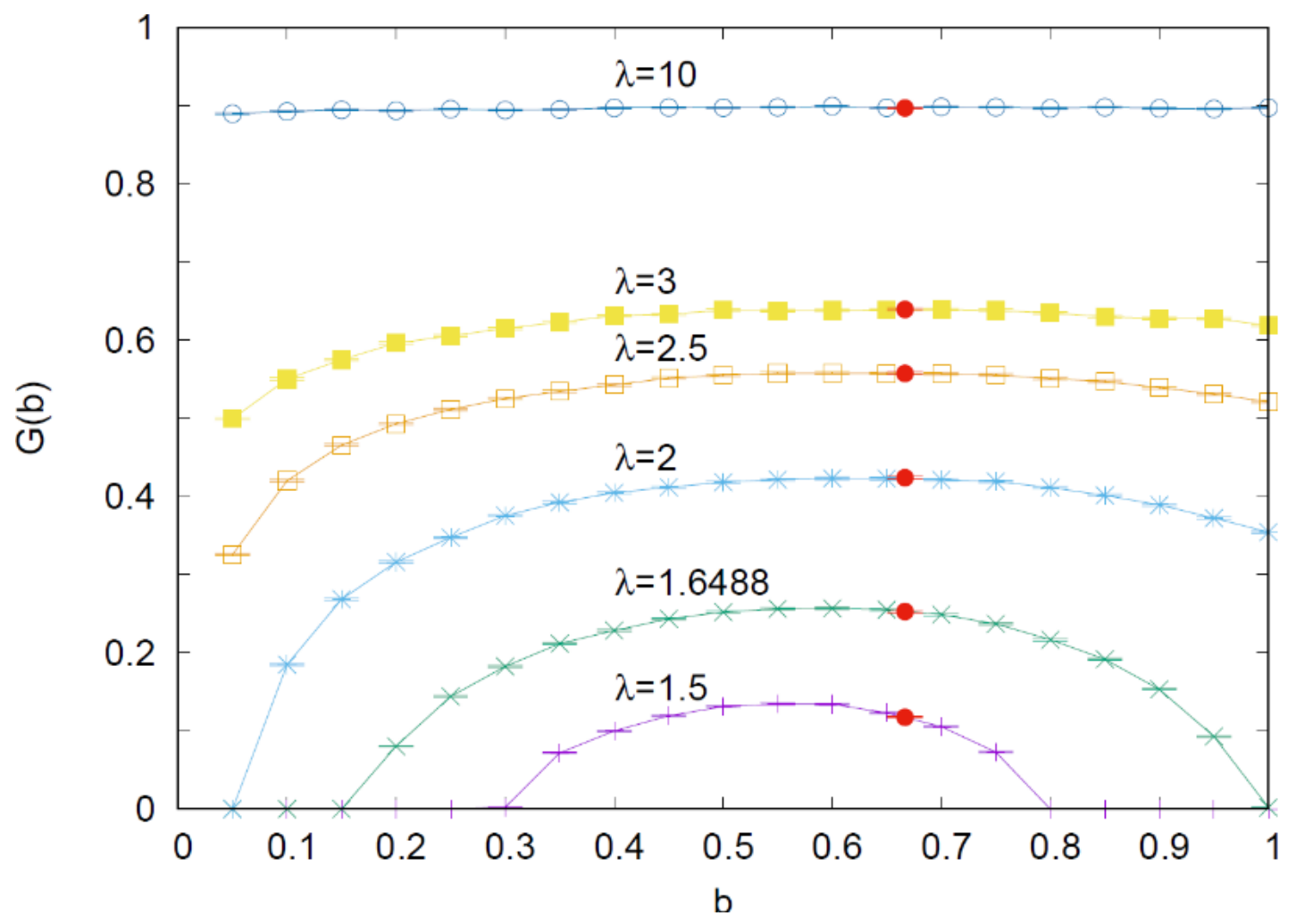

For example, the light blue curve in **Figure 5** shows a case in which the system was only slightly above the standard CP's survival threshold ($\lambda = 2$, while $\lambda_c \cong 1.6448$), as the birth rate of ramets was only twice as high as their death rate. Nevertheless, the genet survived over a broad range of side-branching probabilities, $0.05 \leq b \leq 1$. The range of *b* within which the difference in *G(b)* between neutral and non-neutral growth was less than 0.1 included both forward-growing and side-branching plants ($0.25 \leq b \leq 1$). At the lower limit (*b*=0.25), the tendency for forward growth was quite strong: the average length of linear segments was 4 lattice units. The upper limit (*b*=1) represents extremely strong side-branching, with no forward growth. Thus, even relatively low λ permitted similar survival across diverse branching strategies. At a high value of $\lambda$=10, most branching probabilities resulted in similar survival values; only extremely infrequent branching (very low *b*) decreased the survival probability substantially.

## 4. Discussion

Our results suggest that genet survival can strongly depend on the probability of side-branching, but the strength of this effect depends on the ratio of birth to death rates of ramets. The baseline model for our study was a well-known, paradigmatic model of spatial spreading, the standard contact process (section 2.1). First we compare this standard CP to the newly introduced directional contact process (DCP). Then we compare three kinds of DCPs, which differ in their tendency to grow forward vs. sideways.

### *4.1. The effect of turning back*

The standard CP is isotropic in the sense that all the neighboring sites are attempted to be colonized with equal probability (**Figure 2.a**). Conversely, the directional CP did not permit any turn-back from the parent's to the grandparent's site (**Figure 2.b-d**). The effect of this rule could be seen most clearly in the neutral DCP, in which the plant did not have any preference among the non-forbidden sites; that is, they were colonized with equal probabilities (**Figure 2.b**). Thus, directionality was manifested solely in the absence of backtracking. This proved to be sufficient to significantly enhance survival, lowering the survival threshold (compare the red dot with the dashed line in **Figure 4**). This advantage of the neutral DCP arises because the grandparent's site is more likely to be non-colonizable than the other three sites, as the grandparent may have survived, keeping the site occupied.

The $\lambda$ value at which the standard CP already got extinct ($\lambda_c \cong 1.6448$) enabled rather high survival probability for the neutral DCP, in which $G(\lambda = 1.6448) = 0.252(1)$ (see the dot on the green curve in **Figure 5**). In general, the $G(\lambda)$ function was monotonically increasing with $\lambda$ at any value of *b*. This can be explained by the fact that higher $\lambda$ causes higher occupancy over the area (Oborny et al., 2007). At the same time, the difference between the standard and directional CPs diminished, as the disadvantage of the grandparent's site decreased relative to the other neighboring sites.

The standard CP can be considered a paradigmatic model of spreading without any preferred direction. A good example of this is seed dispersal. Even though external forces, like wind, can impose a direction on dispersal, this direction is global over the whole area (or a considerable part of it, compared to the ramet size), and, more importantly, is not actively influenced by the plant. In contrast, the direction of clonal spreading is influenced by the growth rules of the plant (see Oborny et al., 2012 for a review). In general, our model highlights a significant advantage of reproduction by clonal growth over reproduction by seeds: the directional nature of clonal growth can significantly reduce the frequency of backtracking, thus decreasing competition between generations.

### *4.2. To keep on growing or to change direction?*

This "plant dilemma" was expressed in the DCP model as a trade-off between forward and sideways growth (**Figure 2.b-d**). We found that the genet's survival threshold ($\lambda_c$) strongly depended on the probability of side-branching (*b*). The lowest survival threshold, and hence the broadest range of survival, occurred in moderately forward-growing plants, in which the average length of linear (non-branching) intervals was 1.89.

Above the threshold, the *G(b)* relationship strongly depended on λ. Genet survival was promoted by forward growth in sparse ramet populations (low λ). However, this advantage gradually disappeared as λ increased, and the branching probability became effectively neutral across a broad parameter range encompassing both forward-growing and side-branching plants. This suggests that natural selection for low branching intensity is strong in unfavorable habitats, but weak or negligible in favorable ones.

We think that the explanation lies in a twofold limitation of growth: by the availability of empty sites, and by the availability of meristems. The advantage of forward growth (low $b$) is that it allows the plant to escape from densely populated regions to reach vacant regions. Such vacant regions are typically available at the edges of growing clones during the transient stage or at low values of $\lambda$ (i.e., in unfavourable habitats) in the steady state. Alternatively, high $b$ creates more growing tips, increasing the potential for more ramet production. Its downside is that it increases the probability of branch encounters (**Figure 1**). Optimal allocation between forward growth and side branching hinges on the balance between these advantages and disadvantages. A ubiquitous result was that very low $b$ was always disadvantageous, regardless of the value of λ, due to its severe limitation of meristem availability. Approaching one-dimensional growth naturally increases the survival threshold (which is $\lambda_c \cong 3.306(2)$ in one dimension); that is, it leads to extinction over a broader range of $\lambda$.

We assumed homogeneous habitats throughout the study, to focus on the spontaneously forming branching structure, without any external influence. Nevertheless, the results also suggest some insights into clonal growth in heterogeneous habitats. Several studies have examied the responses of clonal plants to environmental heterogeneity (in our terms, spatially varying $\lambda$). These studies have yielded significant results about the phenotypic plasticity of side-branching probability ($b$) in response to the local quality of the environment (local $\lambda$). Several plant species were found to increase the branching intensity in favourable (high $\lambda$) sites (see Hutchings and de Kroon, 1994; Oborny and Cain, 1997). Computer models have shown that this response can significantly increase the occupancy in a broad range of heterogeneous habitat types. Our study confirms that the response is feasible in terms of the resulting branching structure. We found that low branching intensity was more advantageous in unfavourable habitats (at low $\lambda$), while a broad range of branching intensities were tolerated in favourable habitats (at high $\lambda$); thus the aforementioned plastic response is feasible. Note that this argument is not based on the usual considerations of the plant's probability of encountering favourable sites, but solely on the probability of encounters between branches.

### *4.3. Outlook*

Our model is relatively simple, and therefore it can yield only qualitative results (concerning the shapes of the functions). Quantitative predictions would require more complex models, representing particular species. These models may take into account the exact branching pattern and may consider more details of ramet demography (e.g., the age-dependence of ramet birth and death). Nevertheless, some of our qualitative results may also generalize to more complex models. The transition from the living to the extinct state in the standard CP is known to be a universal phenomenon (see 2.4). This suggests that our findings may be broadly generalizable in the vicinity of extinction. On the other hand, some plant properties may qualitatively change the behavior of the system. Stolons may avoid each other vertically, using the third spatial dimension. The inclusion of more plant properties would be an exciting task for future research. We believe that a better understanding of the trade-off between forward and sideways growth would provide important cues for studying the physiological control of branching (e.g. apical dominance), and the evolution of branching patterns. It would also enhance our understanding of intraclonal competition, and consequently, of spatial pattern formation in plant communities.

## Acknowledgements

This work was supported by the Hungarian National Research, Development and Innovation Office NKFIH (Grant No. K146736); and the Czech Science Foundation (Grant No. 26-21009S).